\documentclass{article}

\usepackage{changes}
\usepackage{cancel}
\usepackage{booktabs}
\usepackage[english]{babel}
\usepackage{lipsum}
\usepackage{amsfonts}
\usepackage{dsfont}
\usepackage{graphicx}
\usepackage{epstopdf}
\usepackage{algorithmic}
\usepackage{xcolor}
\usepackage{array}
\usepackage{float}
\usepackage{mathrsfs}
\usepackage{mathtools}
\usepackage{mathdots}
\usepackage{multirow}
\usepackage{amstext}
\usepackage{amsmath}
\usepackage{amssymb}
\usepackage{enumitem}
\usepackage{amsopn}
\usepackage[T1]{fontenc}
\usepackage[latin9]{inputenc}
\usepackage{sansmathaccent}
\usepackage{multirow}
\usepackage{natbib}
\usepackage{arydshln}
\usepackage[title]{appendix}
\usepackage{setspace}
\usepackage{url}

\newtheorem{theorem}{Theorem}

\newenvironment{proof}[1][Proof]{\noindent\textbf{#1.} }{\ \rule{0.5em}{0.5em}}
\usepackage{subcaption}
\usepackage{graphicx}
\usepackage[affil-it]{authblk}

\begin{document}
	\title{The effect of the pandemic on complex socio-economic systems: community detection induced by communicability}
	
	\author[1]{Gian Paolo Clemente \thanks{gianpaolo.clemente@unicatt.it}}
	\author[2]{Rosanna Grassi \thanks{rosanna.grassi@unimib.it}}
	\author[2]{Giorgio Rizzini \thanks{giorgio.rizzini@unimib.it}}
	
	\affil[1]{Universit\`{a}  Cattolica del Sacro Cuore - Dipartimento di Matematica per le Scienze Economiche, Finanziarie ed Attuariali, Via Necchi, 9 - 20123 Milano MI - Italy}
\affil[2]{Universit\`{a} degli Studi di Milano-Bicocca - Dipartimento di Statistica e Metodi Quantitativi, via Bicocca degli Arcimboldi, 8 - 20216 Milano MI - Italy}

\date{}

\maketitle

\begin{abstract}

The increasing complexity of interrelated systems has made the use of multiplex networks an important tool for explaining the nature of relations between elements in the system. 
In this paper, we aim at investigating various aspects of countries' behaviour during the coronavirus pandemic period. By means of a multiplex network we consider simultaneously stringency index values, COVID-19 infections and international trade data, in order to detect clusters of countries that showed a similar reaction to the pandemic.
We propose a new methodological approach based on the Estrada communicability for identifying communities on a multiplex network, based on a two-step optimization. At first, we determine the optimal inter-layer intensity between levels by minimizing a distance function. Hence, the optimal inter-layer intensity is used 
to detect communities on each layer. 
Our findings show that the community detection on this multiplex network has greater information power than classical methods for single-layer networks. Our approach better reveals clusters on each layer with respect to the application of the same approach on each single-layer. Moreover, detected groups in the multiplex case benefit of a higher cohesion, leading to identifying on each layer a lower number of communities with respect to the ones obtained in the single-layer cases.

\end{abstract}


\section{Introduction}

There is a clear relation between COVID-19 disease, containment measures, and socio-economic factors, as pointed out in \citep{paez2020,Sharma2021,sha2020}.\\
The COVID-19 pandemic occurred in a globally interconnected system involving several aspects of the human life. At the beginning of the pandemic, the socio-economic framework contributed to spread the disease.  Indeed, the flow of people travelling around the world or spending time together favoured the virus diffusion (see, e.g., \citep{AFF2021,AlSalem,Fernandez2020}).
Some factors link wealth with the COVID-19 pandemic, such as the age of people in different countries, the international flows of goods and of tourists, the endowment of the health facilities. These aspects reasonably played an important role in spreading the infection. \\
For these reasons, the containment measures adopted by countries had an impact on these contexts. To rapidly face the pandemic policymakers dealt with the trade-off between safeguarding public health and mitigating the negative economic impact of epidemic containment measures (see, e.g.,  \citep{ijerph2020,Ferrari2021,barbero2021,kok2020,baig2021}).\\
These reasons naturally lead to investigate this complex system with network tools, exploiting the mathematical properties and the characteristics of multiplex networks. In this way, we take into account simultaneously different facets of COVID-19 diffusion (see \cite{montes2020} for interesting insights on this issue). To this end, we aim at detecting communities of countries that showed a similar reaction to the pandemic.

Community detection is not an easy task in a complex system. At a global level the system can appear as a dense and compact set of relations, so that the mesoscale structure does not easily emerge.\\ 
In this context, the concept of node distance plays an important role, since nodes distant from a given one can influence it while passing through intermediary ones.
Recently, \cite{Bartesaghi2020} propose an innovative methodology to detect communities based on the Estrada communicability distance \citep{Estrada2009, Estrada2019}. Unless the classical Euclidean distance, the proposed measure also considers indirect connections becoming crucial to catch deep interconnections between nodes.
Starting from this idea, we propose a methodology to detect communities in a multiplex framework describing different aspect of the pandemic, namely the effect of the containment measures, by the stringency index values, the COVID-19 infections data and international trade data. Indeed, classical approaches based on monoplex do not consider that entities can interact simultaneously in many ways. 
In a multiplex, communities have to be affected not only by the topological structure of the network in each layer, but also by inter-layer connections. 
Since the pandemic broke out, some works studied the link between the commercial trades and the COVID-19 diffusion using network tools (see, e.g., \cite{Antonietti2021bis,reissl2021,kiyota2021,fagiolo2020}).
Differently from \cite{AFF2021} that investigate the effect on the socio-economic framework of the first wave of COVID-19 pandemic, our approach allows to analyse the COVID-19 infection taking into account jointly the adoption of diversified containment policies and the international trades.
As a result, we obtain on each layer a lower number of communities ensuring a higher cohesion with respect to the one that would be obtained focusing only on single layers. On COVID-19 and stringency index layers, what emerges is a world split into two part, at each level. In particular, we group together countries in which COVID-19 has been already detected and characterized by a prompt reaction of the policymakers. 

The World Trade Network (WTN) layer is featured by a strong persistence in the mesoscale structure. Indeed, we argue that the pandemic effect of the first wave did not have an impact on medium-long run connections between countries, strengthening the geographical relationships.

The paper is organized as follows. In section \ref{PreliminariesCommDist}  we remind some preliminary definitions and the Estrada communicability distance. In section \ref{model} we explain the two-steps optimization procedure. In section \ref{numerics} we construct the multiplex network based on the data, and in subsection \ref{secresults} we discuss the results. Section \ref{conclusions} reports conclusions. Appendices \ref{AppA} and \ref{Appendix_cf} provide some mathematical details of the distance function. In appendix \ref{listcountries} the list of countries is given.

\section{Communicability in multiplex networks}
\label{PreliminariesCommDist}


\subsection{Preliminary Definitions}

A network is formally represented by a graph $G=(V,E),$ where $V$ is the set of $n$ nodes and $E$ is the set of $m$ edges (or links). We say that two nodes are adjacent if an edge $(i,j) \in E$ exists between them. The network is undirected i.e. $(j,i)$ is an element of $E$ whenever $(i,j)$ is such. A $i-j$-path is a sequence of distinct vertices and edges between $i$ and $j$. The shortest path, or geodesic, between $i$ and $j$ is a path with the minimum number of edges. The length of a geodesic is called geodesic distance or shortest path distance $d(i,j)$. A graph $G$ is connected if, $\forall i,j \in V$, a $i-j$-path connecting them exists. The degree $k_i$ of node $i$ is the number of edges incident on it.  \\
We assume that the graph is undirected and connected.
The adjacency relations among all nodes are collected into a symmetric $n \times n$ matrix $\mathbf{A}$, called adjacency matrix. No self-loops are allowed, that is $a_{ii}=0$ for $i=1,\dots,n$. \\
If a positive real number $w_{ij}$ is associated to an edge $(i,j)$, we say that both edge and graph are weighted. 
The weighted relations among all nodes are collected into a $n \times n$ matrix $\mathbf{W}$, called weighted adjacency matrix. For further details, the reader can refer to \cite{Estrada2012book}.

Let us now introduce the definition of multiplex network (see \cite{Estrada2014} and \cite{cozzo2018}).
Formally, a multiplex network $M$ (multiplex, for short) is defined by a family of networks $G_{\alpha} = (V,E_{\alpha})$, $\alpha = 1,\dots,h$ and a family of sets of links $E_{\alpha, \beta}$, with $\alpha, \beta = 1,\dots,h$ and  $\alpha \neq \beta$. \\
Each network $G_{\alpha} = (V,E_{\alpha})$ is located into a layer $\alpha$ of the multiplex. 

\begin{center}
\begin{figure}
	\includegraphics[scale=0.5]{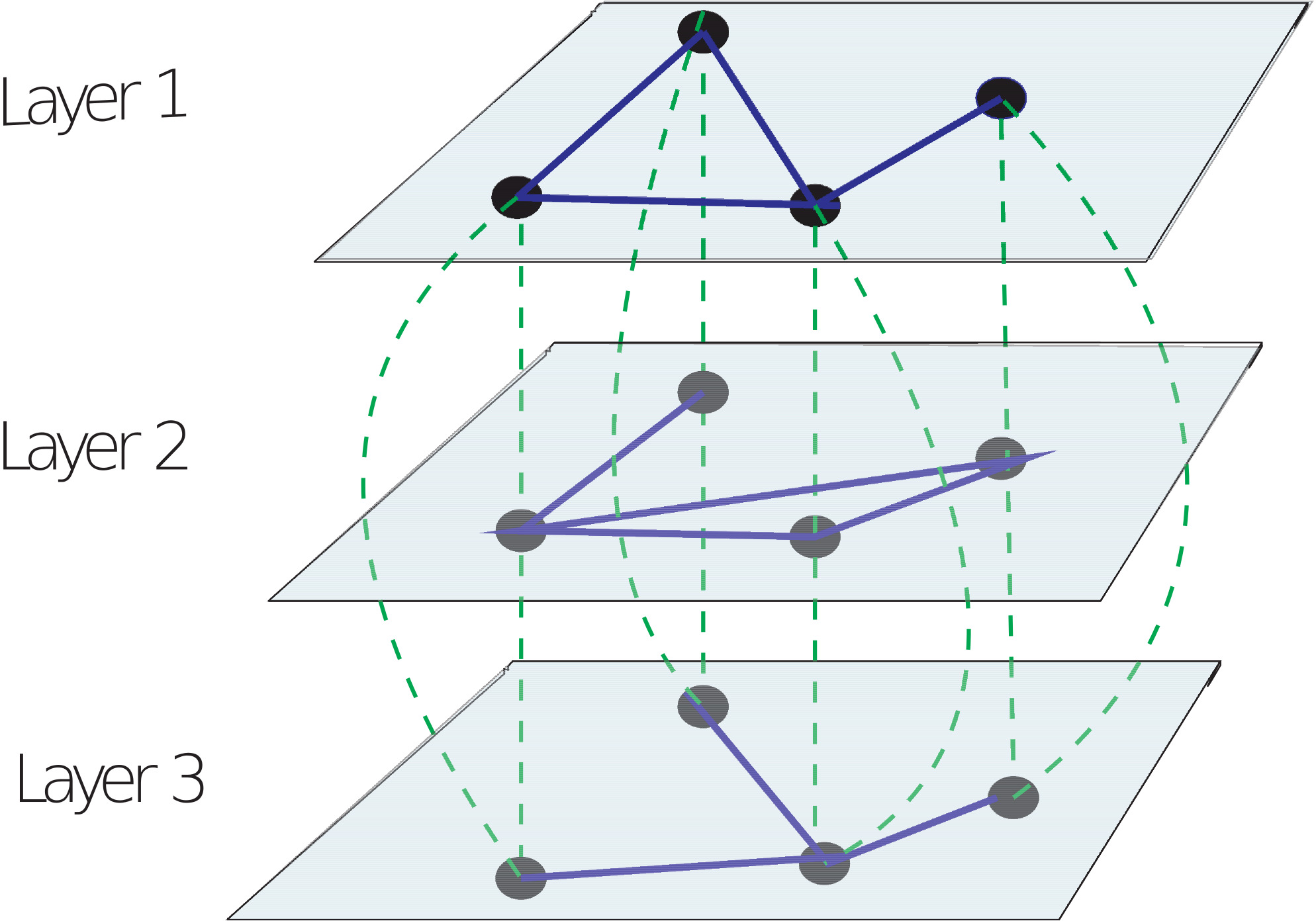}
	\caption{Multiplex network with three layers. Blue links are the intra-layer connections. Green dashed links are the inter-layer connections.}
	\label{figure_three_layers}
\end{figure}
\end{center}

Two nodes $i$ and $j$, $i\neq j$ are connected by a link if they are located in the same level (intra-layer connection), that is $(i,j) \in E_{\alpha}$. Conversely, a node $i$ can be linked only with itself in different layers (inter-layer connection), that is $(i,i) \in E_{\alpha, \beta}$.
In other words, intra-layer connections are the links lying on the same layer, while the inter-layer connections refer to links connecting two layers. A graphical representation of a multiplex is provided in Figure \ref{figure_three_layers}.\\
The topology of the multiplex network is completely described by the supra-adjacency matrix $\mathcal{A}$ defined by 
\begin{align*}
\mathcal{A}
&  = 	\begin{pmatrix}
\mathbf{A}_{1} & \mathbf{0} & \dots & \mathbf{0}\\
\mathbf{0} & \mathbf{A}_{2} & \dots & \mathbf{0}\\
\vdots & \vdots & \ddots & \vdots\\
\mathbf{0} &  \mathbf{0} & \dots & \mathbf{A}_{h}
\end{pmatrix}
+
\begin{pmatrix}
0 & \mathbf{C}_{12} & \dots & \mathbf{C}_{1h}\\
\mathbf{C}_{21} & 0 & \dots & \mathbf{C}_{2h}\\
\vdots & \vdots & \ddots & 0\\
\mathbf{C}_{h1} &  \mathbf{C}_{h2} & \dots &  0\\
\end{pmatrix}\\
&= \mathbf{A}_L + \mathbf{C}_{LL}\\
\end{align*}
being $\mathbf{A}_L$ the block matrix of the adjacency matrices $\mathbf{A}_\alpha, \alpha=1,...,h$ while $\mathbf{C}_{LL}$ expresses the block matrix of all the inter-layer connections. 
Each element of $\mathbf{C}_{LL}$ is a matrix representing the connections between corresponding nodes in different layers. In this case, we assume that the multiplex is made by $h$ layers. \\
Analogously, we can define a weighted supradjacency matrix $\mathcal{W}=\mathbf{W}_L + \mathbf{C}_{LL}$, where $\mathbf{W}_L$ is the block matrix of the weighted adjacency matrices $\mathbf{W}_\alpha, \alpha=1,...,h.$

As standing hypotheses, we assume that weighted adjacency matrices $\mathbf{W}_\alpha$ have weights $w_{ij,\alpha} \in \left[0,1\right]$ and $\mathbf{C}_{\alpha \beta}=\mathbf{C}_{\beta \alpha} =\omega \mathbf{I}$, where $\omega$ is a positive real parameter that expresses the intensity of inter-layer connections and $\mathbf{I}$ is the $n$-square identity matrix. 
Note that the intensity $\omega$ expresses the strength of the connection between two layers. Low (high) values of $\omega$ indicate a weak (strong) connection between them.\\
In particular, $\omega=0$ corresponds to the case in which all layers are independent, that is each layer cannot influence the other ones. In this case, the multiplex becomes a set of single layer networks, not interrelated each others. On the contrary, when $\omega$ is increasing, the weight between different nodes in the same level are neglected respect to the intensity of the same node in different levels. To the scope of our study, the latter case is not significant, being interested in evaluating both types of connections.
Hence, we assume that 
$\omega \in [0,1]$ to be consistent with the fact that  $w_{ij,\alpha} \in \left[0,1\right]$.

\subsection{Estrada communicability distance}
We are interested in analysing how a pair of nodes communicate in a multiplex. To this end, we can use the definition of network communicability. The main idea has been introduced by  \cite{Estrada2012} and then extended to multiplex networks in \cite{Estrada2014}. \\
Formally, the communicability between nodes $i$ and $j$ of a multiplex $M$ is defined by 
\begin{equation}
G_{ij} = \sum_{k=0}^{\infty} \frac{[\mathcal{A}^k]_{ij}}{k!} = \big[\exp(\mathcal{A})\big]_{ij}.
\label{comm_estrada}
\end{equation}
The quantity $G_{ij}$ counts for the number of walks of length $k$ that connect nodes $i$ and $j$ in the multiplex, giving more weight to the shortest routes connecting them. We will refer to this definition as Estrada communicability. \\
It is worth noting that in the multiplex case, the Estrada communicability $G_{ij}$ between nodes $i$ and $j$ does not consider only the intra-layer connections, as in the case of monoplex networks, but also the inter-layer connections. The latter crucially depend on how layers are connected, hence, at the end, on the parameter $\omega$. 
As in the monoplex case, the quantity $G_{ii}$ expresses the importance of the node $i$ in the multiplex and it is known in the literature as subgraph centrality \citep{Estrada2005}.\\
In the weighted case, following \cite{Crofts2009} and \cite{Estrada2012book} we normalize the matrix $\mathcal{W}$ as $\mathcal{S}^{-\frac{1}{2}} \mathcal{W} \mathcal{S}^{-\frac{1}{2}}$, where $\mathcal{S}$ is a $nh$-square diagonal matrix containing the strengths of the nodes for each layer.
Then we define the Estrada communicability as:

\begin{equation}
G_{ij} = \sum_{k=0}^{\infty} \frac{1}{k!} \bigg[\bigg( \mathcal{S}^{-\frac{1}{2}} \mathcal{W} \mathcal{S}^{-\frac{1}{2}}\bigg)^k\bigg]_{ij} = \bigg[e^{\mathcal{S}^{-\frac{1}{2}} \mathcal{W} \mathcal{S}^{-\frac{1}{2}}}\bigg]_{ij}
\end{equation}

The communicability matrix $\mathcal{G}$ is a block matrix defined as 
\begin{equation}
\mathcal{G} =\exp(\mathcal{A})=  
\begin{pmatrix}
\mathbf{G}_1 & \mathbf{G}_{12} & \dots & \mathbf{G}_{1h}\\
\mathbf{G}_{21} & \mathbf{G}_2 & \dots & \mathbf{G}_{2h}\\
\vdots & \vdots & \ddots & \vdots\\
\mathbf{G}_{h1} & \mathbf{G}_{h2} & \dots &  \mathbf{G}_h \\
\end{pmatrix}.
\label{matrix_comm}
\end{equation}
Matrix $\mathbf{G}_{\alpha \beta},\alpha,\beta=1,\dots,h$ is an $n$-square matrix representing the communicability between a node in the layer $\alpha$ and itself in layer $\beta$, while $\mathbf{G}_{\alpha}$ expresses the communicability between every pair of nodes in layer $\alpha$.\\

Fixing two nodes, $i$ and $j$, it is possible to define a distance between them using the Estrada communicability in Equation \eqref{comm_estrada}. The communicability distance $\xi_{ij}$ between any nodes $i$ and $j$ in a multiplex  \citep{Estrada2019} is defined by
\begin{equation}
\xi_{ij} = G_{ii} + G_{jj} - 2 G_{ij}.
\label{comm_distance}
\end{equation}

This formula defines a metric on the multiplex network then it measures how two nodes are distant in communicating, wherever they are located. 
Note that $G_{ii}$ and $G_{jj}$ is the subgraph centrality of nodes $i$ and $j$, respectively, while $G_{ij}$ is the Estrada communicability between those nodes, in whatever layer they are. Formula \eqref{comm_distance} accounts for the information flow passing between $i$ and $j$, net of what each node retains for itself.

\section{Optimization procedure}
\label{model}

In this section we describe the model that allows to obtain the optimal community partition on a multiplex network. As a first step, we 
determine the value of the intensity of the inter-layer connections $\omega^*$ that ensures the minimum total distance between nodes. 
As a second step, using the optimal value $\omega^*$, we compute the Estrada communicability in the multiplex and, later, on each layer, we detect communities based on the communicability distance, defined in formula \eqref{comm_distance}. In the next subsections we describe in detail such methodology.

\subsection{Optimal inter-layer connections intensity}
The optimal inter-layer connections  intensity $\omega^*$ is obtained as a result of an optimization problem. The objective function is a suitable function representing the total distance between nodes in the multiplex network. 
In particular, we define the average distance for a node $i$ and for the network, respectively, as:
\begin{equation}
\bar{\xi}_{i}(\omega)=\frac{1}{nh-1}\sum_{k\neq i}\xi_{ik}(\omega),
\label{xi_bar_i}
\end{equation}
and 
\begin{equation}
\bar{\xi}(\omega)=\frac{1}{nh}\sum_{i}\bar{\xi}_{i}(\omega)=\frac{1}{nh\left( nh-1\right) }\sum_{ij}\xi_{ij}(\omega),
\label{xi_bar}
\end{equation}
where $nh$ is the total number of nodes in the multiplex. Note that the average in formula \eqref{xi_bar_i} is computed dividing the numerator by $nh-1$ because it accounts for all the distances from the node $i$ to the remaining nodes.\\
We then define the multiplex total distance $\Delta_M$ as follows:
\begin{equation}
\Delta_M (\omega) = \sum_i \bar{\xi}_i (\omega) =  nh\bar{\xi} (\omega)\text{,}
\label{coes_multiplex_cf}
\end{equation}



Formula (\ref{coes_multiplex_cf}) defines a global distance in the multiplex as proportional to the average distance for the whole network. This definition is in line with the results obtained for the monoplex network in \cite{Bartesaghi2020} in case of a partition made up of the entire network. In the Appendix  \ref{AppA} we report another expression of the multiplex total distance. 


The following properties hold for the multiplex total distance $\Delta_M$ (we refer the reader to the Appendix \ref{Appendix_cf} for the proof):

\begin{theorem}
\label{AnalyticSol}

\emph{Let $\Delta_M: [0,1] \rightarrow \mathbb{R}$ such that $\Delta_M(\omega) =  nh\bar{\xi} (\omega)$. 
The following properties hold:
\begin{enumerate}
	\item $\Delta_M(\omega)$ is positive 
	\item $\Delta_M(\omega)$ is continuous and differentiable 
	\item $\Delta_M(\omega)$ has global minimum and maximum. If the extreme point $\omega^* \in (0,1)$, the following holds: 
	$$nh\sum_{i}G_{ii}^{\prime }(\omega^*)=\sum_{i,j}G_{ij}^{\prime }(\omega^*).%
	$$
\end{enumerate}
}
\end{theorem}

%
%

Multiplex total distance is strongly affected by the intensity of the inter-layer connections $\omega$ through the quantity $\bar{\xi}$.
Since $\xi_{ij}$ measures how two nodes are distant in communicating in the same and in different layers, we aim at searching the optimal value $\omega^*$ that minimizes the total distance in the network. 


In light of this reasoning, the optimal inter-layer intensity $\omega^{\ast}$ is determined by
\begin{equation}
\omega^* \in \arg \min \Delta_M(\omega).
\label{Optimal_sol}
\end{equation}
The optimal value $\omega^*$ expresses the optimal intensity between layers on the basis of the distance between nodes in terms of communicability. In this way we characterize the multiplex network in a manner that takes into account of the "optimal" communication between its elements.


\subsection{Community Detection}
\label{commDet_layer}
In this section we detect communities on each layer of the multiplex. \\ Specifically, we compute the Estrada communicability matrix  $\mathcal{G}(\omega^*)$ evaluated in the optimal value $\omega^*$ obtained as in equation \eqref{Optimal_sol}.
Then, for each layer, we follow the approach proposed in \cite{Bartesaghi2020} for detect communities in the monoplex case, but 
in our case the Estrada communicability function of each layer contains not only the intra-layer walks but also the inter-layer ones. 

In this way the effect of the inter-layer connections intensity $\omega$ between layers allows us to quantitatively measure the relative position of a node in each layer, taking into account this in detecting communities. The method is briefly discussed below.

\begin{figure}
\centering
\begin{subfigure}[t]{0.3\textwidth}
\centering
\includegraphics[scale=0.30]{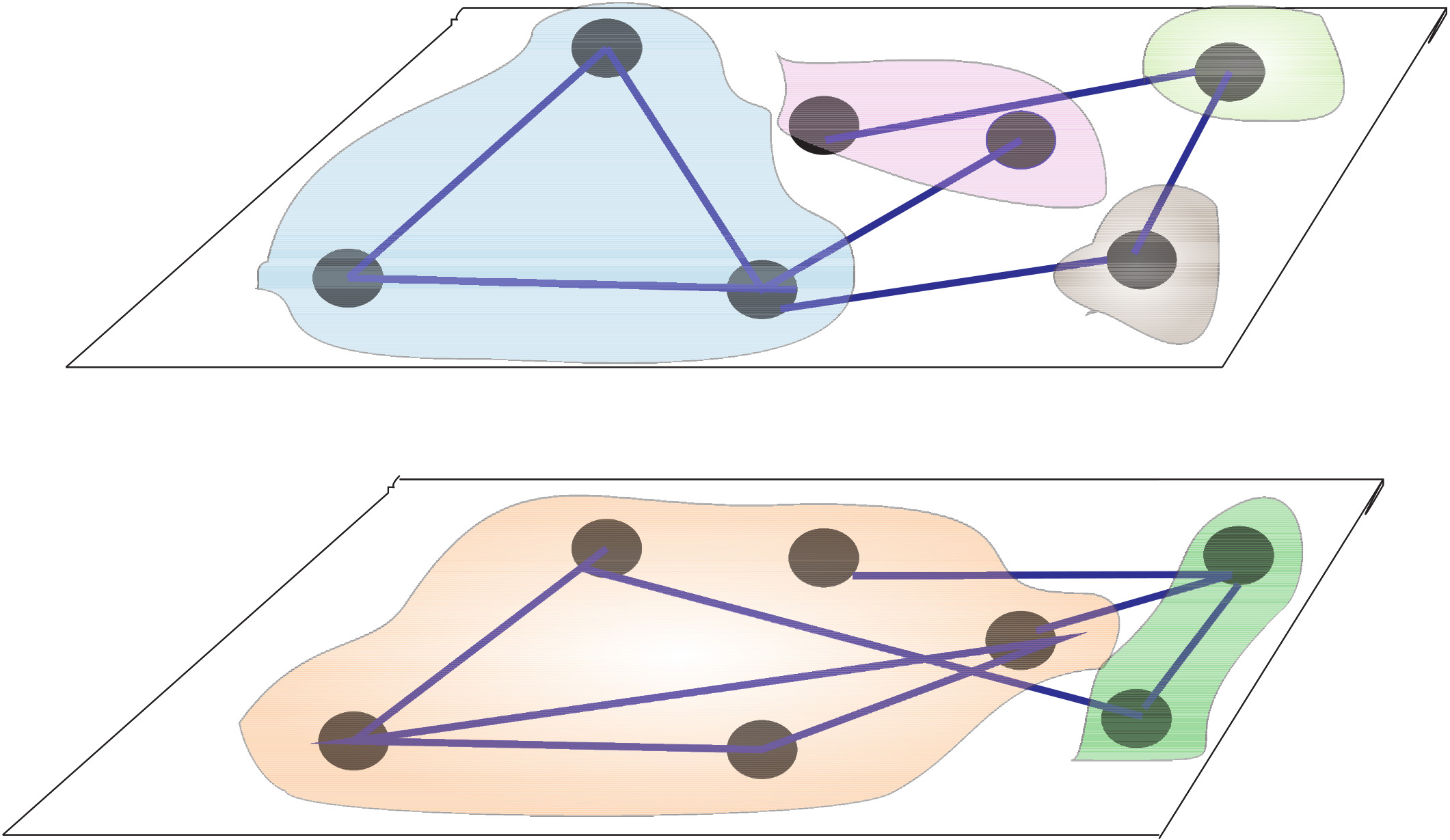}
\caption{Community detection for independent layers ($\omega=0$).}
\label{fig:y equals x}
\end{subfigure}
\hfill
\begin{subfigure}[t]{0.5\textwidth}
\centering
\includegraphics[scale = 0.30]{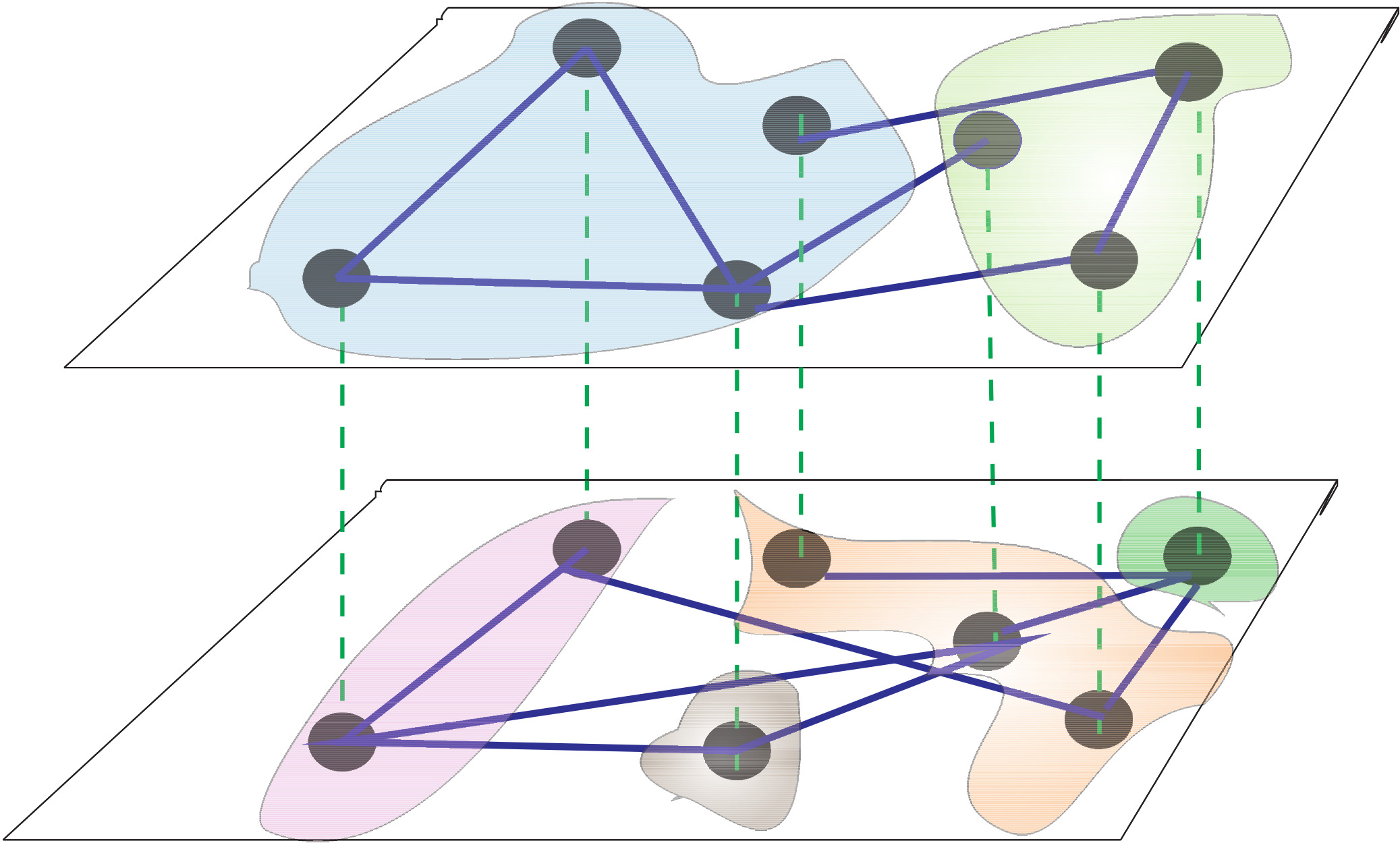}
\caption{Community detection in a multiplex framework ($\omega>0$)}
\label{fig:three sin x}
\end{subfigure}
\caption{Role of $\omega$ in community detection for multiplex networks.}
\label{esempio_semplice}
\end{figure}

For each layer $\alpha$, we consider members of the same community those nodes whose communicability distance is below a given threshold $\xi_{\alpha} \in [\xi_{\min,\alpha}, \xi_{\max,\alpha}]$ which is specific for each layer. The underlying idea is that the more two nodes in a layer are connected, the lower is their communicability distance. Based on  $\xi_{\alpha}$, we construct 
an adjacency matrix $\textbf{M}_{\alpha}=[m_{ij,\alpha}]$ as follows:
$$
m_{ij,\alpha}=
\left\{ 
\begin{array}{ll}
1 & \ {\rm if}\ \xi_{ij,\alpha}\leq \xi_{\alpha}\\ 
0 & \ {\rm otherwise} \\ 
\end{array}
\right.
$$
The optimal threshold $\xi_{\alpha}^*$ is obtained maximizing a quality function $Q_{\alpha}$ 
\begin{equation}\label{Qh}
Q_{\alpha}=\sum_{(i,j)\in E_{\alpha}}\gamma_{ij,\alpha}(\omega^*) x_{ij}
\end{equation}
where $\gamma_{ij,\alpha}(\omega^*)$ is a measure of cohesion (see \cite{chang2016}) between nodes $i$ and $j$ of layer $\alpha$ as 
\begin{equation}
\gamma_{ij,\alpha}(\omega^*)=\left(\bar{\xi}_{j,\alpha}(\omega^*)-\bar{\xi}_{\alpha}(\omega^*)\right)-\left(\xi_{ij,\alpha}(\omega^*)-\bar{\xi}_{i,\alpha}(\omega^*)\right).
\label{gamma}
\end{equation}
and $x_{ij}$ is a variable equal to $1$ if two nodes are in the same community and $0$ otherwise. This allows us to group nodes that are close each others but far by the rest of the network, revealing the clusters of nodes that strongly communicate. Through the maximization of $Q_{\alpha}$ we optimize the division in nodes clusters.
We stress that this approach is in line with those in \cite{Newman2004}. Indeed we do not choose any a priori number of communities, but this number comes out from the maximization procedure. 
Community detection in multiplex network depends strongly on the intensity of inter-layer connections. An illustrative example of how parameter $\omega$ influences communities is shown for two-layers multiplex in Figure \ref{esempio_semplice}.

Concluding, we summarize the proposed methodology in the following steps: 
\begin{enumerate}
\item First step: Multiplex Optimization
\begin{enumerate}
\item compute the multiplex communicability matrix $\mathcal{G}(\omega)$;
\item for all $i$ and $j$ compute the communicability distance $\xi_{ij}$, and their average values, based on the communicability matrix $\mathcal{G}(\omega)$;  
\item compute the closed form of the multiplex total distance $\Delta_M(\omega)$; 
\item find the inter-layer intensity $\omega^*$ such that the total distance $\Delta_M(\omega)$ is minimized.
\end{enumerate}
\item Second step: Layer Optimizations
\begin{enumerate}
\item compute the optimal multiplex communicability $\mathcal{G}(\omega^*)$;
\item for each layer iterate the following steps.
\begin{enumerate}		
	\item for all $i$ and $j$ of the same layer $\alpha$ compute the communicability distance $\xi_{ij, \alpha}$ based on the communicability matrix $\mathcal{G}(\omega^*)$; 
	\item calculate the layer cohesion measure $Q_{\alpha}$;
	\item determine $\xi_{\alpha}^*$ maximizing $Q_{\alpha}$ and find communities that satisfy the inequality $\xi_{ij,\alpha} \le \xi_{\alpha}^*$;
\end{enumerate}
\end{enumerate}
\end{enumerate}

\section{Numerical Section}
\label{numerics}
We provide here an application of the proposed methodology to a multiplex network that considers different aspects of countries' behaviour during the coronavirus pandemic period, also known as COVID-19 pandemic. 

\subsection{Data sources and multiplex network}
\begin{figure}[H]
\begin{center}

\includegraphics[width=12cm,height=6cm]{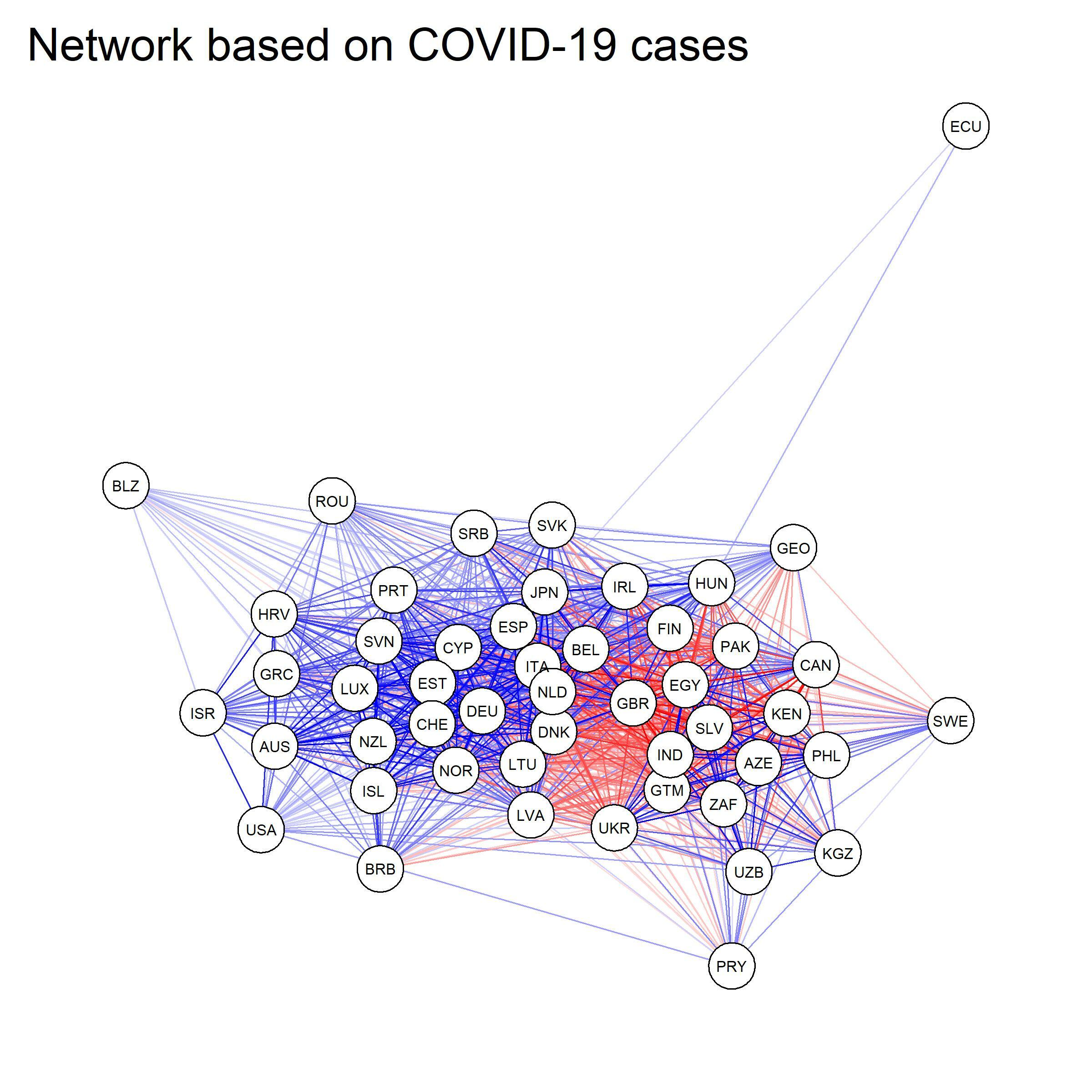}
\includegraphics[width=12cm,height=6cm]{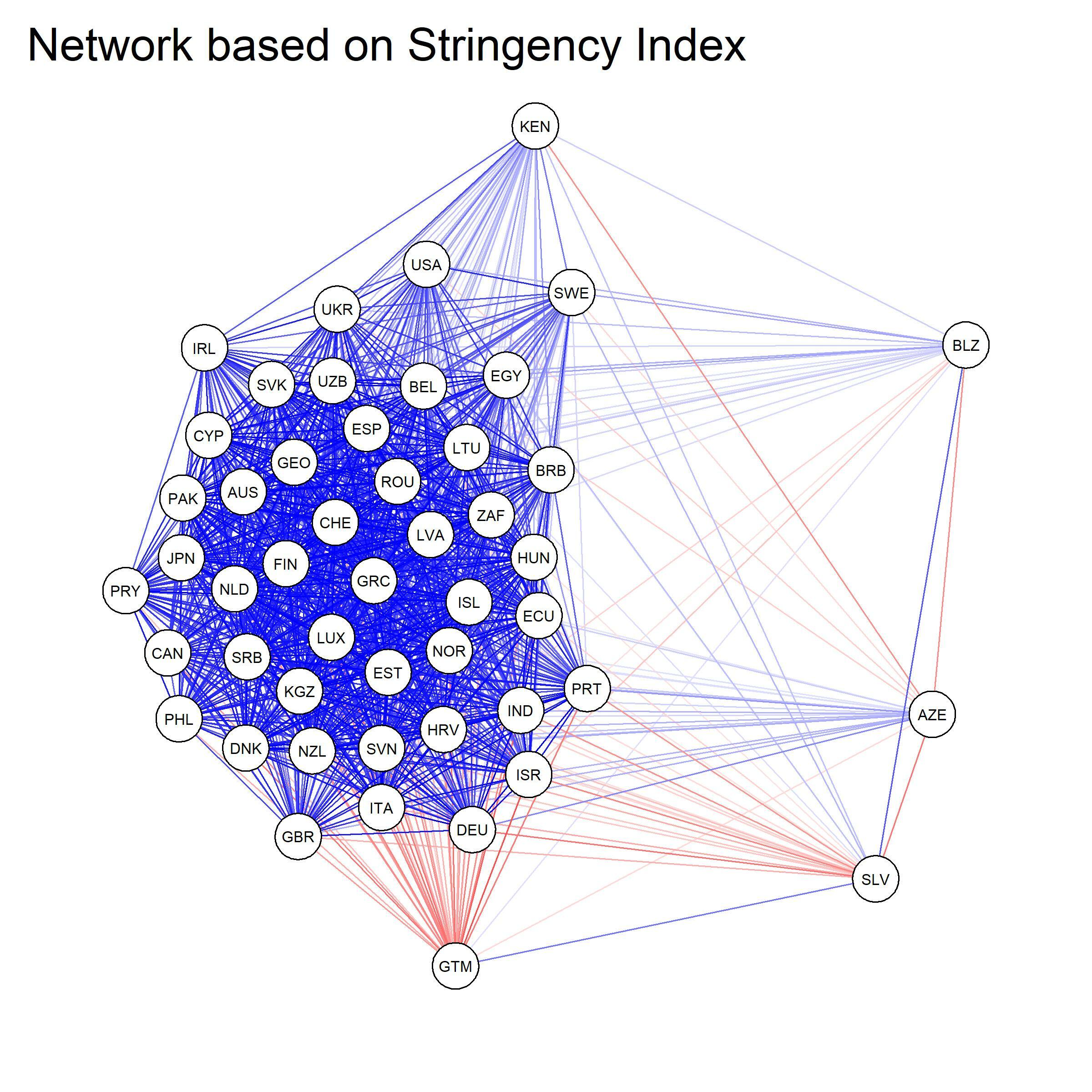}
\includegraphics[width=12cm,height=6cm]{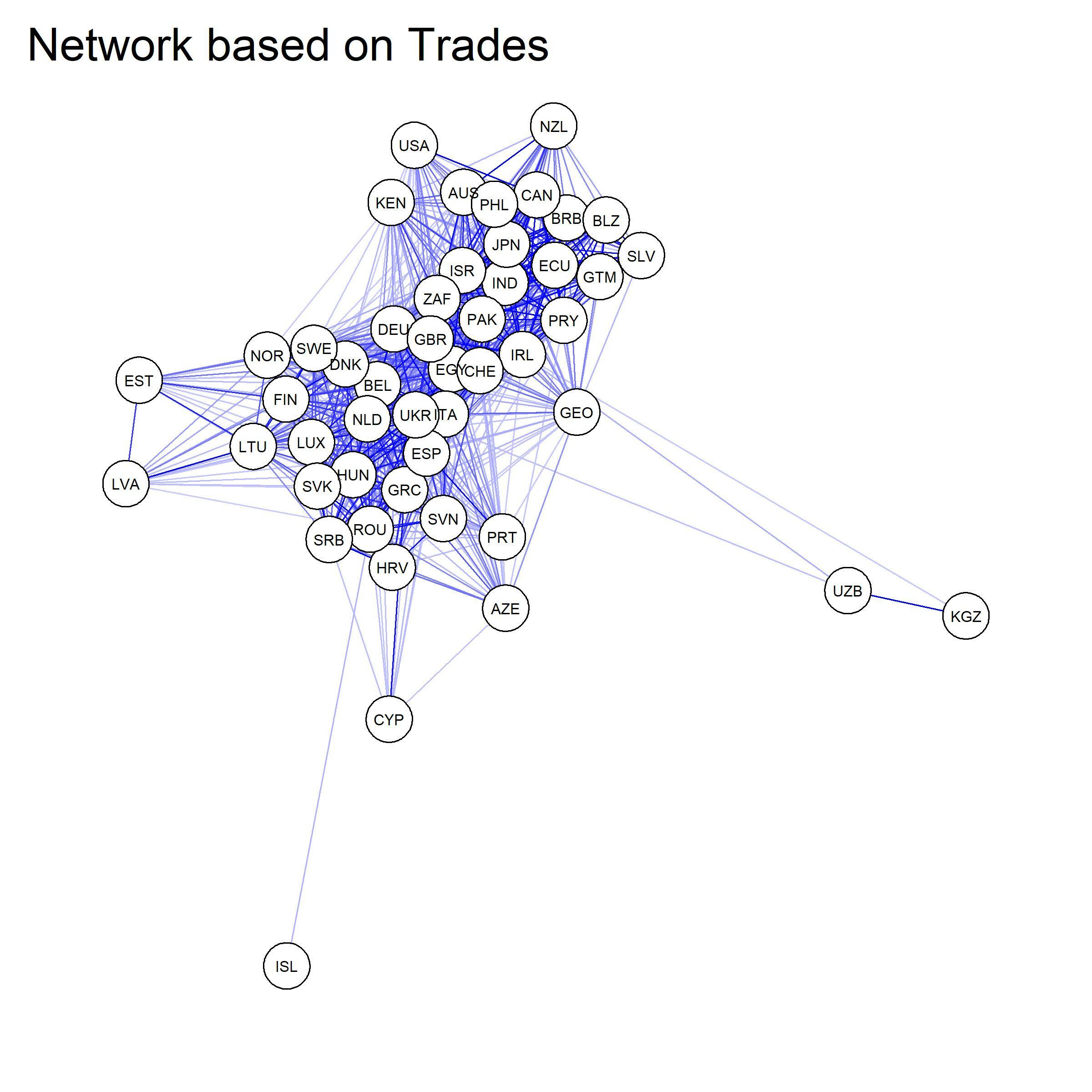}
\caption{Figures display the three single layer networks built assuming a weighted adjacency matrix equal to $\mathbf{D}_{{\alpha}}$, with ${\alpha}=1,2,3$. Opacity of the links is proportional to the absolute value of the correlation coefficient. We report in blue distances related to positive correlations and in red distances related to negative correlations}
\label{fig:Layers}	
\end{center}
\end{figure}

In order to construct the multiplex network used for the numerical application, we make use of three different data sources, two of them strictly related to COVID-19 diffusion. \\ In particular, we focused on the period April 1 - June 30, 2020, that represented a period characterized by both a high diffusion 
of the coronavirus disease and an implementation of heterogeneous response measures between countries. Unfortunately, in the previous period (January-March 2020), also characterized by the diffusion of the disease and by the introduction of initial restrictions, too missing data were present at the worldwide level to assure a good calibration of the network. \\
We collected from \lq\lq Our World in Data\rq\rq, a project of Global Change Data Lab that involves University of Oxford, the daily new cases of COVID-19 disease and the stringency values for each country on daily basis\footnote{Notice that data are also available on github at the following link https://github.com/owid/covid-19-data/tree/master/public/data}. As regards the confirmed cases, the data available come from COVID-19 Data Repository by the Center for Systems Science and Engineering at Johns Hopkins University.
The Government Response Stringency Index (SI) is instead a composite indicator of the government response strategy
to the pandemic diffusion (see \cite{Hale}). This composite measure is a simple additive score of nine indicators evaluated on an ordinal scale and properly rescaled to vary from 0 to 100. Publicly available information on indicators of government response are collected. These indicators measure the level of restriction based on different policies, as school closures, travel bans, etc.\footnote{The nine indicators are school closures, workplace closures, cancel public events, restrictions on gatherings, closure of public transport, public information campaigns, imposing to stay at home, restrictions on internal movement, international travel controls.}. \\
For the same period, we collected the amount of trades between countries from the UN Comtrade, a repository of official international trade data, \cite{UnCOMTRADE}.
To consider the presence of partial missing data in the time series and to build a multiplex network, we compute for each index (e.g. COVID-19 cases) the correlation coefficient between each couple of countries, while we exclude countries for which the whole time series is not available for at least one index. Subsequently, we considered only correlation coefficients that are significant with a $p$-value threshold of 5\%. We define $\rho^{{\alpha}}_{i,j}$ the significant correlation coefficients between countries $i$ and $j$ for the index ${\alpha}$, with ${\alpha}=1,2,3$ representing COVID-19 cases, SI and trade volumes, respectively.\\
In order to assure that the weights of the links range in the interval $[0,1]$, a meaningful solution has been proposed in \cite{Mantegna} (and adapted in \cite{GIUDICI2020}) based on distances
$d^{{\alpha}}_{i,j}=1-\frac{1}{2}\sqrt{2(1-\rho^{{\alpha}}_{i,j})}$. 
The distance matrix $\mathbf{D}_{{\alpha}}=[d^{{\alpha}}_{i,j}]$,
with elements $0\leq d^{{\alpha}}_{i,j}\leq1$, can be interpreted as a weighted adjacency matrix of a single layer network. \\
Hence, we can construct the three distinct a single layer networks $G_{{\alpha}}$ reported in Figure \ref{fig:Layers}. Each network has 49 nodes, that represent countries\footnote{The list of countries and related codes is reported in the Appendix \ref{listcountries}.  We notice that the list of countries does not contain France, Russia and China, which are among the largest in terms of trade. The reason lies in the lack of WTN data from \citep{UnCOMTRADE} for these countries for the second quarter of 2020.}, and a link between nodes $i$ and $j$ is present in the network $G_{{\alpha}}$ only if a significant correlation based on the index $\alpha$ exists between countries $i$ and $j$. In this case, the link has a weight equal to $d^{{\alpha}}_{i,j}$. \\

It is interesting to notice that the network based on COVID-19 cases (see top figure in Figure \ref{fig:Layers}) is characterized by a dense group, with higher average weights, involving mainly European countries and a second group including Eurasian, Asian, Central American and African countries. The two groups are connected by links with very low distances associated to negative correlation coefficients. This is mainly explained by the different diffusion of the pandemic infections during the first wave in these two groups. As shown in \cite{AlSalem}, Italy, Belgium, France, Germany, Great Britain, Netherland, Spain, showed the highest number of cases in Europe in the analysed period with very high peaks of infections and deaths in April 2020. These country show indeed a very high correlation each other. The second group is instead represented by countries that was affected lately by the diffusion. \\
It is also interesting to note the different behaviour of Sweden, that seems not so closed to the other European countries providing a negative correlation. Sweden indeed has been characterized by a different national strategy with respect to the other countries and it has represented an outlier with  cases  and  deaths  increasing  more  rapidly  
than in its Nordic neighbours in the second wave of COVID-19 (see, e.g., \cite{Claeson} and \cite{GORDON2021}). It is also noticeable the behaviour of Japan, United States and Israel that seem connected to the dense European group but at the same time, to confirm their specific behaviour. 
Finally, this first network is very dense, with a density equal to 0.85, confirming that the correlation coefficients
were significant in the larger part of cases. \\
The second network in Figure \ref{fig:Layers} is based on the values of the SI and therefore, it is related to alternative strategies introduced by governments in the analysed period. First of all, we observe a higher average distance with respect to the other networks. This means that the correlation between countries is significantly higher in this case\footnote{The average correlation in the network based on COVID-19 cases is 0.15. It is equal to 0.71 for the SI index and 0.60 for the world trade.}. Although every country's lockdown is different and a wide range of measures have been adopted by national governments, a high correlation between Stringency Index is observed on average. This result shows on average homogenous indices between countries although characterized by combinations of different restriction policies.
Also Sweden shows a positive correlation with other countries but provides the lowest average value in Europe. This is mainly motivated by the fact that the country had relaxed requirement during the fist pandemic wave. For instance, Sweden maintained an open border in the first half of 2020 within the EU and beyond and imposed no quarantine requirements for arrivals (\cite{GORDON2021}). In addition to an open border, Sweden minimised its government-mandated social distancing measures and placed a greater reliance on voluntary behaviours of Swedish residents to comply with national health advisories (see, e.g., \cite{Paterlini} and \cite{Ludvig}). \\
Specific patterns can be observed also for some American countries (as Guatemala and El Salvador), characterized by a initial lower incidence of the virus and by low level of restrictions in that period.\\
The third network in Figure \ref{fig:Layers} is related to volume of trades exchanged between countries. This network shows a lower density (equal to 0.63) with respect to the other networks because of a high presence of non-significant correlations. On the other hand, all the correlation coefficients are positive showing similar economic dynamics between countries in the period. This evidence can be justified by a common reduction trend observed between countries. Indeed, worldwide merchandise trade flows decreased significantly in 2020, as COVID-19 disrupted economic activity across the globe. Various pandemic-related factors shaped international trade flows, specifically, COVID-19 incidence and lockdown restrictions affected the volume of imported and exported goods. For instance, \cite{Liu} find that government measures to curb economic activities had a larger impact on China's imports than the direct health and behavioural effects of the pandemic itself. \\

Given the strict relations between the three dynamics involved, we construct a multiplex network based on a supradjacency matrix $\mathcal{W}$  built as described in the methodological section. In particular, we consider
$\mathbf{W}_L$ the block matrix of the adjacency matrices $\mathbf{D}^{\alpha}, \alpha=1,...,3$ of the layers and the block matrix $\mathbf{C}_{LL}$ of all the inter-layer connections based on the parameter $\omega$ that expresses the constant intensity of inter-layer connections.

\subsection{Results}
\label{secresults}

In order to test the proposed methodology, the community detection approach has been initially applied separately on each network $G_{\alpha}$. It is noteworthy that this is equivalent to the application of the proposed method with $\omega=0$.
Then, the proposed methodology has been computed according to the steps described in Section \ref{model} starting from the whole multiplex network (i.e. $\omega > 0$).

In the multiplex case the community detection process is based on two steps. Firstly, the optimal inter-layer intensity $\omega^*$ has been assessed via formula \eqref{coes_multiplex_cf} and subsequently, the community detection method provided in Section \ref{commDet_layer} has been applied. The aim is to emphasize possible differences with respect to the single layer case and the advantages of a multiplex framework.\\
A first interesting evidence is that the application of the methodology considering the whole network $M$ assured for each layer a higher cohesion $Q_{\alpha}$ with respect to the separate computation of the same approach on each layer. Therefore, clusters detected in the multiplex case benefit of a higher cohesion, leading to identify on each layer a lower number of communities as displayed in Table \ref{tab:clust}.

\begin{table}[H]
\begin{center}
\begin{tabular}	{|l|c|c|c|}
	\hline
	& \multicolumn{3}{c|}{Number of clusters} \\
	\hline		
	& COVID-19 & Stringency Index & Trade  \\ 
	\hline
	Single layer & 10 & 28 & 17 \\
	Multiplex & 6 & 12 & 8 \\
	\hline
\end{tabular}
\caption{Clusters detected by applying the proposed methodology on the single layer and multiplex networks, respectively.}
\label{tab:clust}
\end{center}
\end{table}

\subsubsection{Daily cases of COVID-19 contagion}
\begin{figure}[H]
\begin{center}

\includegraphics[width=12cm,height=7cm]{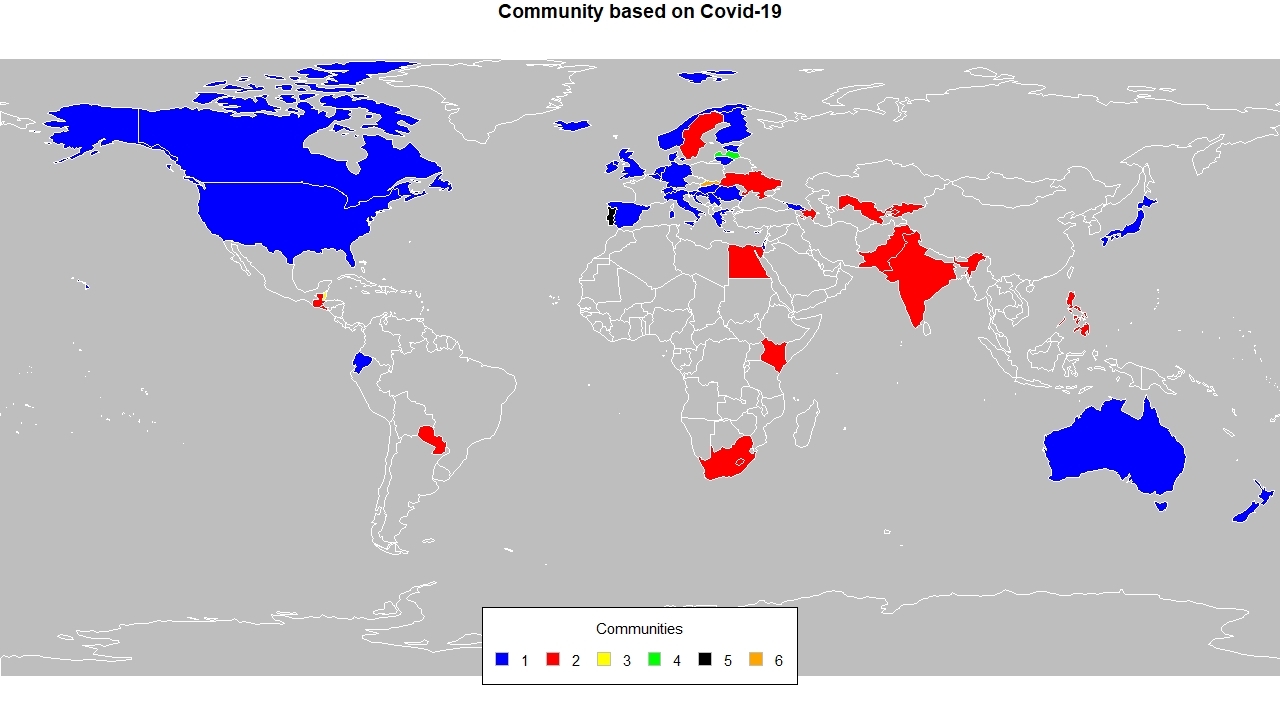}
\caption{Communities detected with the multiplex approach for the network based on Covid-19}
\label{fig:CommCov}	
\end{center}
\end{figure}

Focusing on the clusters obtained with the proposed methodology for the COVID-19 network, we observe two relevant communities in Figure \ref{fig:CommCov}. The larger one collects together most part of European countries, United Kingdom, Unites States and Australia. This result could be explained by the evidence of a similar pattern of all these countries in terms of disease diffusion in the analysed period. The pattern of the pandemic has indeed varied across countries, but, for instance, a common feature across Europe has been the presence of a first wave occurring in the spring of 2020. This wave has been followed by a consistent reduction in the spread of the virus in the summer and a resumption of the epidemic activity in 2020 with second and/or third waves (see, e.g., \cite{Remuzzi}). In the first phase Italy was the original epicentre of the COVID-19 pandemic in Europe, with a subsequent spread in Western and Northern Europe (see, e.g., \cite{Villanie006422}). United Kingdom, Spain, Germany, countries of Benelux Union, Denmark, Norway and Finland are indeed characterized by similar levels of cases in the period and  belong to the same community (see community 1 in Figure \ref{fig:CommCov}). Eastern Europe was instead less affected by the virus in the first phase. For instance, Slovakia (that in Figure \ref{fig:CommCov} is not included in community 1) was less affected by the COVID-19 virus in the first wave and has shown an increase in cases, and consequently in deaths, in the second phase (see \cite{Villanie006422}). \\
A second larger community includes Asian, Eurasian, Central American and African countries. Therefore, countries characterized by a lower incidence in the fist wave. Also Sweden is present in this group being characterized by an anomalous behaviour in that period with respect to other nordic regions (\cite{Claeson}). \\

\subsubsection{Stringency Index}

\begin{figure}[H]
\begin{center}

\includegraphics[width=12cm,height=7cm]{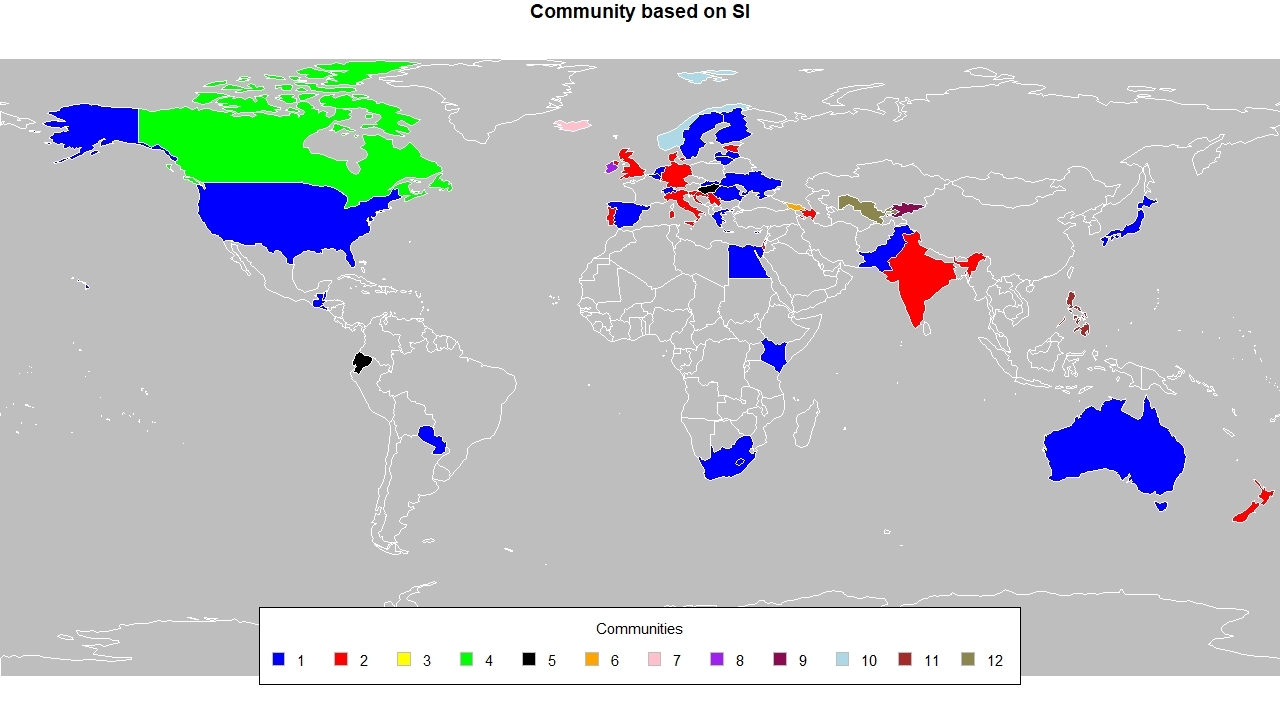}
\caption{Communities detected with the multiplex approach for the network based on the Stringency Index}
\label{fig:CommSI}	
\end{center}
\end{figure}

A larger heterogeneity has been instead observed for the Stringency Index network with twelve communities identified by the procedure (see Figure \ref{fig:CommSI}). However, also in this case, two big communities are detected. Indeed, the pandemic has varied in different countries or regions due to both differences in the capacity of countries to adapt their health system to the COVID-19 epidemic and in governmental policy responses. \\
Certainly, governments around the globe have been taking a wide range of social control measures in response to the COVID-19 outbreak with different degrees and intensities. For instance, as shown in \cite{Ma2021}, countries have been characterized by a different first day of response (i.e. first day with Stringency Index higher than zero) to the situation. Few Asian countries reacted in the first half of January, while European countries, United States and Canada introduced the first measures between the second half of January and the month of February. Other European countries (as Sweden and several states of Eastern Europe), Brazil and Mexico introduced the first measures only at the end of February and during the month of March. Finally, the rest of the world (mainly Kazakistan, Chile, Bolivia, and several African countries) initiated a response after $11^{\text{th}}$ of March 2020, the day World Health Organization (WHO) declared COVID-19 as a global pandemic. \\
However, our data focus on the period April-June 2020, hence, the results appear more affected by the average level of restrictions in the period. In particular, countries characterized by very high level of Stringency Index are on average in the community 2 (see Figure \ref{fig:CommSI}). To this group belongs countries that upgraded the SI to a high level and kept it for a while. In particular, Italy, Central European countries (Netherlands, Luxembourg, Germany, Denmark), countries that was belonging to the former Yugoslavia (Slovenia, Serbia, Croatia), Portugal, Israel, India are all characterized by values of the Stringency Index that belong to the top quartiles of the Stringency Index distribution in the analysis made by \cite{Ma2021}. On the other hand, community 1 in Figure \ref{fig:CommSI} includes countries that maintained lower or moderate levels of the Stringency Index as Sweden, United States, Canada, Australia, and Japan.

\subsubsection{World Trade Network}
\begin{figure}[H]
\begin{center}

\includegraphics[width=12cm,height=7cm]{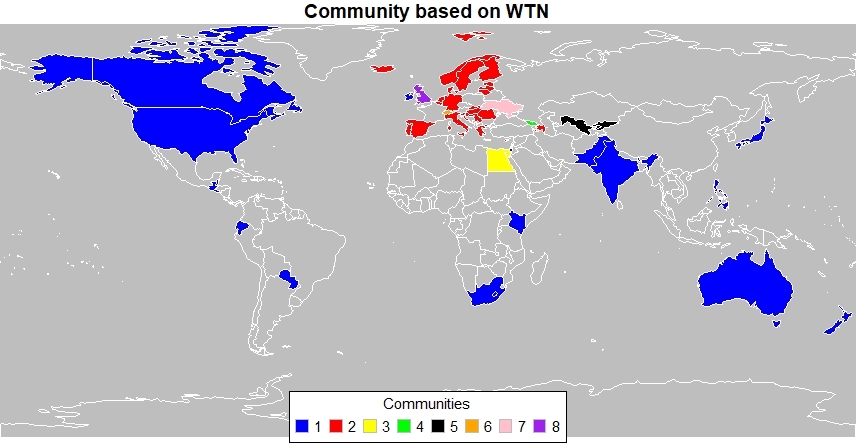}
\caption{Communities detected with the multiplex approach for the network based on world trade}
\label{fig:CommWTN}	
\end{center}
\end{figure}

As regards the economic trades, on the one hand we notice a large Asian-American-Oceanian community and on the other hand, a European cluster. It seems that the pattern of trade during the pandemic period strengthened the geographical relationships between countries. In particular, in Europe there are preferential channels of internal exchanges, whereas, outside Europe, most communication
channels seem to be polarized around the exchanges between American and Asian countries.
The health crisis indeed leaded to confused and non-cooperative responses, characterized by more or less direct and formal export restrictions, including within the European Union (EU) (see \cite{Bald} and \cite{Bown}). In particular, as described in \cite{Evenett}, countries responded to the COVID-19 pandemic with various combinations of export controls and import liberalisation measures. Some countries acted on both sides creating long-term changes in their pre-COVID-19 trade policy structures for the food and medical sectors. Other countries ruled only on one side, either restricting trade or liberalising it. Although EU Member States reacted with a certain level of heterogeneity, a common commercial effect is noticeable also due to the role played by the  European Commission. This behaviour can motivate the composition of community 2 in Figure \ref{fig:CommWTN}. \\
The methodology has been also tested considering the WTN during the second quarter of 2019 to highlight possible differences. We have that main results obtained for $2020$ are still valid for $2019$. In particular, it is confirmed also in 2019 the higher cohesion with the multiplex approach, with a reduction of the number of communities. Focusing on clusters, the partition in $2019$ is similar to the one in $2020$. This result is not surprising and it is in line with the findings in \cite{antonietti2021}. The authors show that WTN is indeed characterized by a persistence of the community structure also during the first part of the COVID-19 pandemic. This results could be explained by the fact that aggregated data do not allow to show peculiar effects of specific sectors (such as, for example, pharmaceutical industry). Additionally, the analysis focus only on the first period of the COVID-19 pandemic situation and reasonably, main impact cannot be reflected immediately on the trade volumes.
\\

To detect possible similarities, we compare the clusters obtained on the three layers. Several approaches for comparing clusters have been proposed in the literature (see \cite{Danonsurv} for a review). Here, we apply the normalised mutual information index, a discriminatory measure based on similarity between partitions (see \cite{Amelio}, \cite{Lanci}). The measure ranges in the interval $[0,1]$ assuming the maximum value when identical partitions are detected. We found the maximal value between pairs, equal to 0.35, when trade and SI are considered. Hence, this level of similarity denotes an important relationship between COVID-19 government restrictions and development of trades.  \\

\section{Conclusions}
\label{conclusions}

In this paper, we assess the relationship between COVID-19 infections, containment measures and World Trade Network during the second quarter of 2020 via a complex network framework. The analysis has been performed through a mathematical approach based on a multiplex network with three layers. We highlight the interdependence through an inter-layer intensity parameter. We determine the optimal value that minimizes the average distance of the entire multiplex. Then, we perform a community detection on each layer based on the Estrada communicability distance. Our results show that through the multiplex approach we obtain a higher flow of information which allows to refine the communities obtained. Indeed, we observe that the clusters are characterized by a higher cohesion, due to a lower internal distance. Focusing on COVID-19 layer, two big clusters emerge characterized by a strong presence and a low occurrence of infections, respectively. A similar dichotomy is observed in the stringency index layer, allowing to highlight similarities in the policymaker decisions. Considering World Trade Network, European, Pacific and North European clusters have been obtained. In this case, a persistence of the mesoscale structure is confirmed with a strengthening of the geographical relationships between countries.

\bibliographystyle{chicago}

\bibliography{References}

\begin{appendices}

\section{}
\label{AppA}

We show that the function $\Delta_M(\omega)$ can be rewritten as sum of $\gamma_{ij}$, where $\gamma_{ij}=
\left(\bar{\xi}_{j}-\bar{\xi}\right)-\left(\xi_{ij}-\bar{\xi}_i\right)$ is a measure of cohesion between nodes $i$ and $j$ (see \cite{chang2016}).
Specifically, $\left(\xi_{ij}-\bar{\xi}_i\right)$ represents the
relative distance between nodes $i$ and $j$ and $\left(\bar{\xi}_{j}-\bar{\xi}\right)$ represents the relative distance
from a random node to the node $j$. Two nodes are more cohesive if they are (relatively) more close each other than they are to any other node in the network. 

\begin{eqnarray*}
\Delta _{M}(\omega ) 
&=&nh\bar{\xi}\\
&=&2\left(nh\right)^{2}\bar{\xi}-2\left(nh\right)^{2}\bar{\xi}+nh\bar{\xi}\\
&=&\left( nh\right) ^{2}\bar{\xi}-\left(
nh\right) \left( nh-1\right) \bar{\xi}+\left( nh\right) ^{2}\bar{\xi}-\left(
nh\right) ^{2}\bar{\xi}
\end{eqnarray*}

Recalling that, by formulas \eqref{xi_bar_i} and \eqref{xi_bar}, $\sum_{i}\bar{\xi}_{i}=(nh)\bar{\xi}$ and $\sum_{ij}\xi_{ij}={nh\left( nh-1\right) }\bar{\xi}$ we have:
\begin{eqnarray*}
\Delta _{M}(\omega )&=& nh\left( \sum_{j}\bar{\xi}_{j}\right) -\sum_{ij}\xi _{ij}+nh\left(
\sum_{i}\bar{\xi}_{i}\right) -\left( nh\right) \left( nh\right) \bar{\xi}\\
&=&\sum_{i}\left( \sum_{j}\bar{\xi}_{j}\right) -\sum_{ij}\xi
_{ij}+\sum_{j}\left( \sum_{i}\bar{\xi}_{i}\right) -\sum_{ij}\bar{\xi} \\
&=&\sum_{ij}\bar{\xi}_{j}-\sum_{ij}\bar{\xi}-\sum_{ij}\xi _{ij}+\sum_{ij}	\bar{\xi}_{i} \\
&=&\sum_{ij}\left( \left( \bar{\xi}_{j}-\bar{\xi}\right) -\left( \xi _{ij}-\bar{\xi}_{i}\right) \right) =\sum_{ij}\gamma _{ij}.
\end{eqnarray*}

\section{Proof of Theorem \ref{AnalyticSol}}
\label{Appendix_cf}

We report in this appendix the proof of Theorem 1. \\
We first recall the following results concerning the series of complex-valued functions (see \cite{rudin1964}, Ch. 7), that obviously also hold for real-valued ones:
\begin{theorem}[Test for uniform convergence]
\label{Rudin10}
Let $\{f_n\}$ be a sequence of complex-valued continuous functions on a set $E$ in a metric space and $f$ a complex-valued function defined in $E$. Suppose $|f_n(x)| \le M_n$, $\forall x \in E$. Then $\sum f_n$converges uniformly on $E$ if $\sum M_n$ converges.
\end{theorem}

\begin{theorem}
\label{Rudin12}
Let $\{f_n\}$ be a sequence of complex-valued continuous functions on $E$ and $f$ a complex-valued function defined in $E$. If $\{f_n\}$ converges uniformly on $E$ to the function $f$, then $f$ is continuous on $E$.
\end{theorem}

\begin{theorem}
\label{Rudin17}
Let $\{f_n\}$ be a sequence of complex-valued differentiable functions on $[a,b]$  such that $\{f_n(x_0)\}$ converges for some point $x_0$ on $[a,b]$. If $\{f'_n\}$  converges uniformly on $[a,b]$, then $\{f_n\}$ converges uniformly on $[a,b]$ to a function ${f}$ and $\lim_{n \rightarrow \infty}f'_n(x)=f'(x)$, $x \in [a,b]$.
\end{theorem}

Theorems \eqref{Rudin12} and \eqref{Rudin17} can be easily extended to the series of functions, being a series of functions uniformly convergent if the sequence of partial sums is.

\subsubsection*{Proof of Theorem 1}

\begin{proof}
\begin{enumerate}
	\item $\Delta_M$ is positive, being the product of $nh$ and the average distance $\bar{\xi}$. 
	\item  By formula \eqref{coes_multiplex_cf}:
	\begin{eqnarray*}
		\Delta_M(\omega) &=& nh \bar{\xi}(\omega)\\
		&= & \frac{nh}{nh(nh-1)} \sum_{ij} \xi_{ij}(\omega) \\
		& = & \frac{1}{(nh-1)} \sum_{ij} \xi_{ij}(\omega)
	\end{eqnarray*}
	being 
	$\xi_{ij} = G_{ii} + G_{jj} -2G_{ij}$, where $ G_{ij} =  \sum_{k=0}^{\infty} \frac{[\mathcal{A}^k]_{ij}}{k!}$.\\
	To prove the continuity and differentiation of $\Delta_M(\omega)$, by linearity  it is enough to show that $G_{ij}$ is a continuous and differentiable function.
	
	The $ij$-entries $[\mathcal{A}^k]_{ij}$ are monotonic polynomial functions of at most degree $k$ in the variable $\omega$, we denote these functions by $P_k(\omega)_{ij}$, then $\sum_{k=0}^{\infty} \frac{P_k(\omega)_{ij}}{k!}
	$ is a series of continuous functions. 
	
	As $P_k(\omega)_{ij}$ are monotonic functions in $[0,1]$, it is immediate to see that: 
	
	
	$$
	\frac{P_k(\omega)_{ij}}{k!} \le \frac{P_k(1)_{ij}}{k!} \hspace{2mm} \forall k >0, \hspace{2mm} \forall i,j \in V,
	$$
	and the series $\sum_{k=0}^{\infty} \frac{P_k(1)_{ij}}{k!}$ is convergent.
	
	Then, by Theorem \eqref{Rudin10}, $\sum_{k=0}^{\infty} \frac{P_k(\omega)_{ij}}{k!}
	$ uniformly converges, and, by Theorem \eqref{Rudin12}, $G_{ij}$ is a continuous function.
	
	The $ij$-entries $[\mathcal{A}^k]_{ij}$ are differentiable, with derivatives that are polynomial functions of at most degree $k-1$ in the variable $\omega$. Moreover, the series of derivatives uniformly converges, then by Theorem \eqref{Rudin17}, the sum of the series is $G'_{ij}$. This proves that $G_{ij}$ is differentiable in $[0,1]$.

	\item By Weierstrass theorem, $\Delta_M(\omega)$ has global minimum and maximum. The first derivative can be written as
	\begin{eqnarray*}
		\Delta_M^{\prime }(\omega) &=&\frac{1}{nh-1}\sum_{ij}\xi _{ij}^{\prime }(\omega) \\
		&=&\frac{1}{nh-1}\sum_{ij}\left( G_{ii}^{\prime }+G_{jj}^{^{\prime
		}}-2G_{ij}^{\prime }\right)  \\
		&=&\frac{1}{nh-1}\left( \sum_{ij}G_{ii}^{\prime
		}+\sum_{ij}G_{jj}^{^{\prime }}-2\sum_{ij}G_{ij}^{\prime }\right)  \\
		&=&\frac{1}{nh-1}\left( nh\sum_{i}G_{ii}^{\prime
		}+nh\sum_{j}G_{jj}^{^{\prime }}-2\sum_{ij}G_{ij}^{\prime }\right)  \\
		&=&\frac{1}{nh-1}\left( nh\sum_{i}G_{ii}^{\prime
		}+nh\sum_{i}G_{ii}^{^{\prime }}-2\sum_{ij}G_{ij}^{\prime }\right)  \\
		&=&\frac{2nh}{nh-1}\sum_{i} G_{ii}^{\prime }-\frac{2}{nh-1}%
		\sum_{ij}G_{ij}^{\prime }\text{.}
	\end{eqnarray*}
	At the extreme points $\omega^* \in (0,1)$, $\Delta'_M(\omega^*) = 0$ then
	$$
	\frac{2nh}{nh-1}\sum_{i}G_{ii}^{\prime }-\frac{2}{nh-1}\sum_{i,j}G_{ij}^{%
		\prime }=0
	$$
	that yields
	$$
	nh\sum_{i}G_{ii}^{\prime }=\sum_{ij}G_{ij}^{\prime }%
	\text{.}
	$$
	
\end{enumerate}

\end{proof}

\section{List of countries}
\label{listcountries}

\begin{table}[H]

\begin{minipage}{0.5\textwidth}
	\begin{tabular}{lc}
		\toprule
		\textbf{Country} & \textbf{Code} \\
		\midrule
		Australia	&	AUS	\\
		Azerbaijan	&	AZE	\\
		Barbados	&	BRB	\\
		Belgium	&	BEL	\\
		Belize	&	BLZ	\\
		Canada	&	CAN	\\
		Croatia	&	HRV	\\
		Cyprus	&	CYP	\\
		Denmark	&	DNK	\\
		Ecuador	&	ECU	\\
		Egypt	&	EGY	\\
		El Salvador	&	SLV	\\
		Estonia	&	EST	\\
		Finland	&	FIN	\\
		Georgia	&	GEO	\\
		Germany	&	DEU	\\
		Greece	&	GRC	\\
		Guatemala	&	GTM	\\
		Hungary	&	HUN	\\
		Iceland	&	ISL	\\
		India	&	IND	\\
		Ireland	&	IRL	\\
		Israel	&	ISR	\\
		Italy	&	ITA	\\
		Japan	&	JPN	\\
		\bottomrule
	\end{tabular}
	
\end{minipage} \hfill
\begin{minipage}{0.5\textwidth}
	\begin{tabular}{lc}
		\toprule
		\textbf{Country} & \textbf{Code}  \\
		\midrule
		Kenya	&	KEN	\\
		Kyrgyzstan	&	KGZ	\\
		Latvia	&	LVA	\\
		Lithuania	&	LTU	\\
		Luxembourg	&	LUX	\\
		Netherlands	&	NLD	\\
		New Zealand	&	NZL	\\
		Norway	&	NOR	\\
		Pakistan	&	PAK	\\
		Paraguay	&	PRY	\\
		Philippines	&	PHL	\\
		Portugal	&	PRT	\\
		Romania	&	ROU	\\
		Republic of Serbia	&	SRB	\\
		Slovakia	&	SVK	\\
		Slovenia	&	SVN	\\
		South Africa	&	ZAF	\\
		Spain	&	ESP	\\
		Sweden	&	SWE	\\
		Switzerland	&	CHE	\\
		Ukraine	&	UKR	\\
		United Kingdom	&	GBR	\\
		United States	&	USA	\\
		Uzbekistan	&	UZB	\\
		\bottomrule
	\end{tabular}
	
\end{minipage}
\end{table}
\end{appendices}

\end{document}